# Hierarchy of Interstellar and Stellar Structures and the Case of the Orion Star-Forming Region

Bruce G. Elmegreen[1] and Yuri N. Efremov[2]

**Abstract.** Hierarchical structure from stellar clusters, to subgroups, to associations and star complexes is discussed in the context of the Orion stellar grouping and its origin. The analogous structure in gas clouds is also reviewed, with an emphasis on general fractal properties and mass distribution functions. The distinction between the cloud, cluster, and stellar mass functions is discussed in terms of different sampling statistics for hierarchically structured clumps in clouds with an interclump medium.

## 1. The history of stellar associations

### 1.1. First recognition

The Orion constellation illustrates how young stars have a tendency to be gathered together. Today we know this is a common property of all young stars, but in fact such clustering was recognized long before any stars were known to be young. Stellar clusters like Orion provided the first real evidence that star formation occurs today, because most of them are unbound and dispersing rather quickly. The same clustering property also tells us something more, that the gas is hierarchically structured, perhaps because of turbulence. Here we review this hierarchical structure; a previous review was in Scalo (1985).

The tendency for OB stars to form groups was known soon after the appearance of the first large catalogues of stellar spectra and proper motions. During 1910 - 1914, Jacobus Kapteyn, Arthur Eddington and Anton Pannekoek found large groupings of these stars, which are now called OB associations, in Orion, Scorpius and Centaurus. Their large densities in the sky plane and their similar proper motions demonstrated that they are real. By 1929, the total number of known OB star groups was 37, owing mainly to the work of Pannekoek (1929).

Some open star clusters (i.e., gravitationally bound clusters) were also found to contain high luminosity stars, and what is especially important, some of them appeared "to be closely connected with a rather widely scattered cloud of bright stars in its neighborhood" (Shapley & Sawyer 1927). Shapley & Sawyer noted that clusters of this kind include NGC 6231, $h$ and $\chi$ Persei, the Orion Nebula group, and probably M11. They concluded "these clusters appear, in fact, to be merely nuclear concentrations in much large organizations".

[1] IBM Research Division, T.J. Watson Research Center, P.O. Box 218, Yorktown Heights, NY 10598 USA, bge@watson.ibm.com

[2] Sternberg Astronomical Institute, Moscow 119899, Russia, efremov@sai.msu.su

The same conclusion was reached by Bidelman (1943), who studied the double cluster *h* and *chi* Per. He found that around both clusters there are about five dozen supergiants with the same distances and velocities. The size of the whole group is about 200 pc, which is much larger than the usual size of an open cluster. Bidelman realized that these supergiants cannot be runaway stars from the double cluster because the cluster is tightly bound by gravity. He noted in conclusion that the problem of the dynamics of star clouds is far from solved.

Another important investigation of that time was by Struve (1944), who studied a similar clustering of supergiants around the open cluster NGC 6231 in Scorpius. He stressed that this tendency for clusters to be surrounded by vast groups of high luminosity stars is one of their most important properties. But Struve, like everyone else at the time, failed to realize that these regions are young. The element of time was introduced to the study of clusters only after their peculiar dynamics was discovered.

## 1.2. Early dynamical considerations

In 1947 these two stellar groupings, h and $\chi$ Per and NGC 6231, were used by Ambartsumian (1949) as examples of rarefied clusterings of OB stars for which he proposed the name *stellar associations*. From an analysis of published data, he noted that the space densities of stars around these associations are too low for them to resist tidal forces from the inner part of the Galaxy. In a time on the order of 10 Myrs, they should become elongated in the galactic plane. Such elongation was not observed then (later it was shown to exist for NGC 6231), so he concluded that the associations should *disperse*, which means they should have some expansion velocity from the very beginning.

The dynamical instability of associations was a very important conclusion. They are unstable but still observable, thus they must be *young*. Both Bidelman and Struve could have made this statement, but they did not. Moreover, it was possible to evaluate the expansion age of an association, and it appeared to be about 10 Myrs, similar to the nuclear age of the high luminosity stars. The real triumph for Ambartsumian came when Blaauw (1952) found that proper motions of stars in the small association around $\zeta$ Perseus displayed an expansion velocity of about 10 km/sec. This result was stressed as a confirmation of Ambartsumian's prediction.

The dynamical confirmation of a young age for high luminosity stars was important in those days. There were still doubts because some people thought high luminosity stars were really old, but only appeared young because of accretion of interstellar gas on their surfaces.

However Ambartsumian went further in his conclusions. He suggested, and later insisted, that the origin of the expansion of an association is the formation of stars by a burst of some superdense unobservable body. Even in 1985 there were publications with this idea.

Today, the superdense explosion hypothesis is not generally believed, but in the 1940s, there was some basis for this conclusion. Ambartsumian argued that gravitationally bound concentrations of diffuse matter can give birth only to gravitationally bound stellar groups. However we know today that the nuclear energy released by hot stars and supernovae is the ultimate source of the cluster's expansion. The stars blow out the gas from the new-born group and the remaining mass is not sufficient to bind the stars together. Modern observations show that only about 10% of the gas is transformed into stars. If an essential part of the initial mass is lost, then a protocluster becomes an expanding association. This idea was proposed by Fritz Zwicky (1953) many years ago, but only during the last decade has it become evident that this mechanism should work (see review by Lada 1991).

Dynamical instability was suggested to be a discriminator between associations and clusters by Lada & Lada (1991). However, recent developments on correlations between the sizes, ages, and velocity dispersions in hierarchically nested clusters suggest that some of the observed expansion of associations might not really be systematic (see Sect. 2.1.). Even the interpretation of the proper motions of the member stars is far from unambiguous (Brown et al. 1997).

### 1.3. East versus West

The nature of stellar associations was a topic of intense debate in the Soviet Union in the beginning of the fifties. This was a time when the Communist Party said that Soviet science should affirm the priority of Russian scientists and battle against "idealistic theories" from Western science. The concept of stellar associations, as declared by Ambartsumian, conformed to these conditions. Indeed, the Ambartsumian model was victorious inside Russia, and the "doctrine of stellar associations", which is the idea that clustered star *formation* is actually occurring in our times, was declared as an "outstanding victory of the Soviet materialist cosmogony". This was the proclamation of the cosmogonic conference held in Moscow in 1952.

Those who talked against the doctrine of stellar associations, with its essential notion that the origin of stars comes from mystic superdense bodies, included Vorontsov-Veliaminov, Lebedinsky, and Gurevich, who were advised "to use more completely the rich actual data that was available". This was a rather mild admonition – indeed, it was still possible to talk against Ambartsumian's ideas even though he had already won in 1950 the prestigious Stalin Prize for "the discovery of a new type of stellar system." This situation in astronomy was not nearly as bad as in biology, where many of those who spoke against Lysenko, (who denied completely the achievements of genetics and promised to increase the harvest), subsequently lost their jobs. In Astronomy the outspoken simply did not gain the most prestigious positions, such as membership in the Academy of Sciences.

The tense atmosphere surrounding research on OB associations made it rather difficult to do original work in any branch of this field. The problem was that the issues of expanding associations, star formation from superdense bodies, and even the very existence of associations as a kind of stellar group, all merged in the minds of many Soviet astronomers. The rule of 'bon ton' was simply to avoid such topics. "Yura, you should realize you only give an advertisement to them," Josef Shklovsky told one of the present authors, having read his popular article *against* the 'doctrine of stellar associations' in the 1970's. Shklovsky later (in 1984) reproached Yu.E. for too soft an evaluation of the doctrine of stellar associations, when (Efremov 1984) wrote about complexes and associations in a prestigious Russian magazine. Indeed, such an article was only possible because of Shklovsky's support. Shklovsky said that the Ambartsumian doctrine is "Lysenkoism," and that the social roots were the same for both.

It is important to realize with hindsight that even at an early time of the "Buracan doctrine" (Ambartsumian was the Director of the Buracan Observatory), there was in fact some understanding of the real situation by Western astronomers. One of them was Otto Struve, the great-grandson of the founder of the Pulkovo Observatory and an American astronomer since his hasty retreat from Russia with the remains of the White Army in 1920.

Struve (1949) wrote an article for Sky & Telescope on stellar associations, stressing the importance of Ambartsumian's conclusion about their dynamical instability. He referred to "the strange and unconventional idea that the stars may not have been produced of dust and gas," yet formed from dark objects of unknown constitution. But in 1952, Struve wrote another paper entitled "Astronomy in the spirit of *1984*" in which he wrote: "Ambartsumian has not 'discovered' the existence of stellar associations, though his big merit is advancing the remarkably stimulating ideas on their properties and origin. Has the memory of Kapteyn disappeared in the Soviet Union? Has the great Dutch astronomer become an 'unperson'?" (Struve 1952).

The assumption of superdense bodies to explain star formation seemed to be so strange that there was no room for it in Western minds. Bart Bok wrote in his remarkable book *Milky Way* about the important works of Ambartsumian on star formation from diffuse matter. One of us (Yu.E.) was the editor of the Russian translation of this book and had to write in the foreword that the real opinion of Ambartsumian on protostellar matter is quite the opposite of what Bok wrote. This statement was necessary because there had been a case before when a technical editor added the words "of gas" after Ambartsumian's sentence on stellar origins from large masses. Masses of what, thought the editor; so he added, naturally, "of gas". There was then a bit of a scandal in Russia, although changing Ambartsumian's statement to one involving dense gas was exactly the right thing to do, considering our current picture of star formation in dense molecular clouds.

### 1.4. Attempts at a modern definition of OB associations

Sometimes the real existence of associations as distinct from open clusters was denied. The Moscow astronomer Pavel Kholopov (1970, 1979, 1981) found, like Shapley and Bidelman much earlier, that all open clusters are surrounded by a vast halo, and he concluded that associations are nothing but young, loose clusters with a halo populated by OB stars. In fact, this is not true because it is more common for associations to exist without bound cluster cores. This is seen very well in images of the Large Magellanic Clouds, where there are many bright stellar groupings without cluster-like concentrations. Nevertheless, the ambiguity in defining clusters and associations as concentric, or *hierarchical*, groupings on different scales, was beginning to be apparent.

Today there is no common definition for an association, although everyone agrees they are larger, younger, and more rarefied than open clusters, and that they should be unbound. There were even suggestions to name associations

"completely unbound stellar groupings" (Lada & Lada 1991). However, the dynamical status of a group is not easy to determine (it requires the total stellar mass and the velocities), and even stellar Complexes (see below), which are much larger than associations like Orion, are unbound.

The usual terminology applied to stellar groups also depends on distance, because the same grouping of luminous stars can look like a compact cluster from a large distance or a rarefied association if it is near-by. In addition, the local associations are detected from the distribution of O-type and early-type B stars (earlier than B2), but in other galaxies, the spectral types are seldom available. Then we are forced to base the definition of an association on our judgement of star positions, and the resulting group can be different in different studies.

Lucke and Hodge (1970) published a catalogue of 122 OB-associations in the LMC. The association sizes were typically 15 - 150 pc (the average was 80 pc). They noted however, that it is often a matter of judgement as to how many separate nuclei should be considered as separate associations, and that there were many more entries in an unpublished listing of LMC associations by Bengt Westerlund.

Some of the Lucke-Hodge associations were much larger than 80 pc, up to 350 pc. These were called "star clouds". Somewhat similar were the groupings detected in the LMC by Harlow Shapley long ago. Shapley (1931) noted 15 subclouds or "small irregular star clouds", "nearly all of which appear to be distinct physical organizations." The sizes of these "subclouds" are in the range of 150 - 400 pc, including 4 NGC objects.

Later on, Shapley (1951) noted that only the smallest groups of supergiants in the LMC are comparable in dimension to what are called galactic clusters: "Many of them have ten times the diameters and luminosities of such galactic clusters as M11 and Pleiades. Should such widespread assemblies be called subclouds or superclusters, or would it not be better to designate them constellations? They are doubtless comparable to the Orion, Scorpio and Vela aggregations of bright galactic stars." Thus the term "constellations" in the LMC appears. A half dozen of them were numbered by McKibben-Nail & Shapley (1953). The best picture is given by van den Bergh (1981).

Even larger were the dimensions of $\sim$ 200 groups of blue stars in M31, some $\sim$ 500 pc in size, which van den Bergh (1964) called OB associations. He believed that their sizes are so large, about 10 times those of associations in the Milky Way, because the denser background in our Galaxy prevents us from seeing the outer, presumably rarefied parts of the local associations. In comparing the results of searches for associations in a number of galaxies, Hodge (1986) concluded that their sizes depend on resolution. With resolution lower than that used by van den Bergh, Hodge found in M31 42 associations with a size of 300 pc. Evidently he was able to find only the brighter parts of van den Bergh's associations, who also noted the existence of bright cores in many of them.

At the same time, Efremov, Ivanov, & Nikolov (1987) used large-scale plates from the 2-m telescope at Rozhen Observatory to make an independent search for stellar associations in M31. They were able to find 203 large groups of blue stars with 650 pc average size. Most of these were more or less identical to the associations found earlier by van den Bergh. Yet by looking for only the brightest blue stars in the U plates, they found 210 smaller groups with an average size of 80 pc, and 90% of these smaller groups were *inside* the larger groups. They concluded that the smaller group are genuine, classical associations, similar to those known in the Milky Way.

The appearance of large groups in M31 is not only the result of a smaller background density. The large groups are mostly older, except for their cores, as determined from the presence of Cepheid variable stars, which are ten times older than O-stars. Every time one of the fields containing van den Bergh's large M31 associations was searched for variable stars, Cepheids with ages up to 50 Myrs were found in concentration within the association (van den Bergh 1964; Efremov 1980).

### 1.5. Why a physical definition for associations is difficult

We shall see in later sections that the difficulty in defining OB associations lies in the nature of the distribution of all young stars, which is actually hierarchical with no characteristic scale. Selection effects contribute to the confusion.

One problem is that, as we know now, there is a correlation between the duration of star formation in a region and its size. This implies that if OB associations are defined to be any close grouping of OB and other *young* stars, say 10 My old or less, then the associations will be observed to have a maximum size of around 100 pc because larger regions will generally contain older stars. There is also another correlation between the mass of the largest star that forms in a region and the cloud mass, resulting in part from statistical sampling of the initial stellar mass function. This second correlation implies there is a minimum likely size for a region with O-type stars.

Thus OB associations, defined to contain O-type types with a maximum age range overall, will always have a characteristic size and associated GMC mass that is similar to what is observed in Orion and other catalogued objects in the local Milky Way. But there is no physical origin for this size and mass, they are only samples from a continuous distribution of properties for stellar clusters and gas. If the maximum age constraint were not imposed, as, for example, in van den Bergh's study of M31, where he searched for groupings of reasonably blue stars without knowing their ages, then the sizes of the objects identified as OB associations can be much larger than the sizes of OB star clusters. That would be like calling all of Gould's Belt an OB association because it is a grouping of OB stars. Indeed from some perspective in another galaxy, without data on the presence of old clusters and supergiants, such a classification might be made. But with an implicit maximum age constraint, Gould's Belt and other similar regions are not called OB associations. Neither are the star-forming regions associated with isolated small clouds, such as Taurus, because these regions typically contain too little gas to be likely to produce a rare O-type star.

Evidently, the term OB association has no physical meaning, only a series of defining constraints that has the effect of selecting, out of a continuous distribution of cluster properties, those in a particular size and mass range. When these constraints are relaxed, stellar groups with other sizes and masses can be called associations. As a result, OB associations looked physically distinct from other stellar groups for a long time in the history of astronomy, until the hierarchical and scale-free nature of interstellar matter and star formation was recognized.

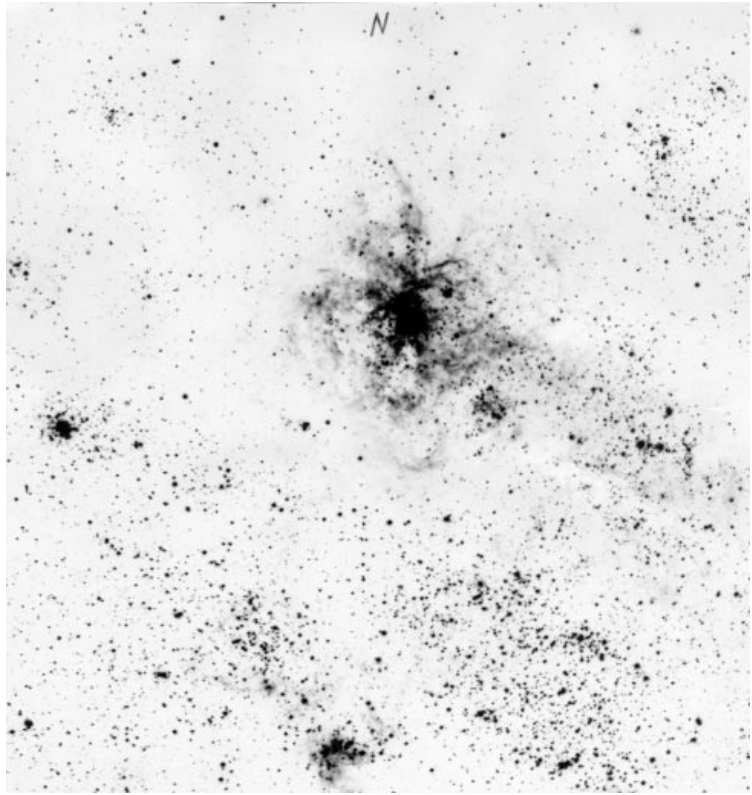

Figure 1. A star field in the LMC centered on 30 Dor, showing several elongated groupings with hierarchical sub-structure. The elongation is probably from shear. This image and Figure 2 were taken with a 70-cm Maksutov camera at Pulkovo Observatory Station in Chile.

The preferred size of 80 pc for OB-associations in the nearest galaxies (Efremov 1995) seems now to be an artifact of the selection of stars of more or less the same age.

## 2. Star complexes and Hierarchical Structure

The M31 stellar groups with $\sim$ 600 pc sizes and $\sim$ 50 My ages turned out to have the same properties as the stellar "complexes" detected in the Milky Way by Efremov (1975, 1978, 1979) using mainly Cepheids. This name of star complexes was suggested for the largest clumped groupings of young stars and clusters in a galaxy. The large associations catalogued by van den Bergh in M31 were similar to star complexes in the Milky Way in every way: age, dimension, stellar content, and tendency to lie along the spiral arms.

It was also recognized that the large, rich complexes detected by Cepheids in the Galaxy are the same as the star clouds and bright knots detected in the spiral arms of other galaxies. Indeed, long ago Seares (1928) believed that these knots are similar to the Local System (the Gould Belt) of rather young stars inside which the Sun is located, and many parameters of this System were quite similar to those observed for more distant complexes. Star complexes are omnipresent in spiral and irregular galaxies, being the largest groupings of young stars inside which associations form and dissolve during the complex's life (Efremov 1979; Elmegreen 1979). Efremov & Sitnik (1988) confirmed that as many as *90% of OB-associations and young clusters in the Galaxy may be united into vast complexes*. The largest complexes detected with Cepheids (Efremov 1978; Berdnikov & Efremov 1989, 1993) are the same as the complexes detected with young clusters and associations.

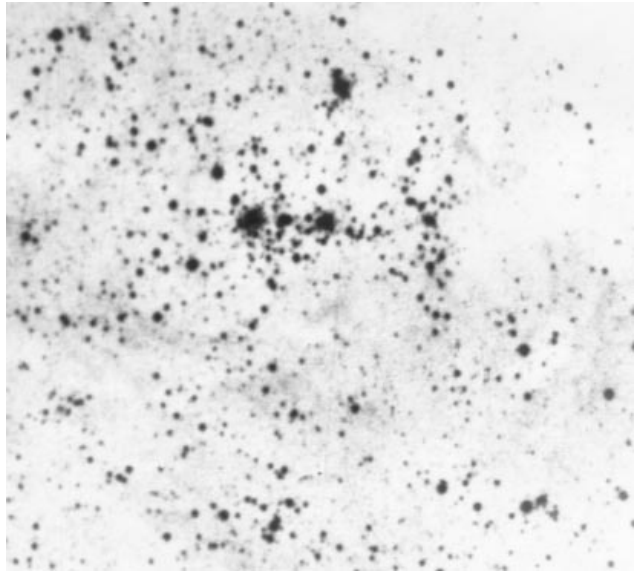

Figure 2. A blow-up of the star field in the previous figure, showing stellar groupings with hierarchical sub-structure southwest of 30 Dor. At the bottom right is the precursor star to SN1987A, seen close to two compact clumps of stars in the tiny cluster KMK80, which is presumably the host of the supernova.

Such hierarchical clustering of young stars was known in the LMC as well. Again something like a star complex was the largest entity. Westerlund & Smith (1964) noted that the distribution of high luminosity stars in the LMC is compatible with the suggestion that star formation is going on in clouds with sizes of 150 - 1000 pc. Then Hodge (1973), having observed that clusters with similar ages form vast groups, concluded that star formation occurs when the gas density is large in areas of 1 kiloparsec size. All subsequent developments on the concept of star complexes confirmed this idea.

The reality of vast groupings of Cepheids, clusters, and supergiants in the LMC was confirmed by Karimova (1989, 1990) with a variety of statistical methods. She found clustering of these objects with scales of about 200 pc and 1400 pc. Earlier, Feitsinger & Braunsfurth (1984) found in the LMC three levels of hierarchical clustering of OB associations and HII regions, with characteristic sizes of 100, 400 and 1500 pc, and Efremov (1984, 1989) noted the hierarchical clustering of associations in the 30 Dor region (figs. 1 and 2). These data may be considered objective indications of the existence of hierarchical sequences of young stellar groupings, from clusters to associations, to aggregates (group of associations), to complexes and regions (supercomplexes). The actual sizes, in parsecs, should not be identified with physically important length scales, however, because the *definitions* for the regions are subjective, as discussed above.

Long ago, in his Harvard lectures of 1958, Baade (1963) stressed that star formation in the LMC occurs on two scales - in associations with dimensions on the order of 10 to 100 pc, and over huge areas with dimensions of 500 pc. The latter scale is that of *superassociations*, which are groups of OB-associations and HII regions. The primary example of a superassociation is the 30 Dor region in the LMC (Baade 1963; Ambartsumian 1964). However superassociations are rare; in the giant spiral M31, only one group was deemed worthy to have this name: NGC 206, the bright star cloud in the southern spiral arm S4 (Baade 1963).

A superassociation seems to be a complex within which all star formation goes on simultaneously over its area (Efremov 1995). The hierarchical inner structure is seen there very well because bright OB stars are numerous in a superassociation. There is never a uniform field of stars but always a number of clusters and associations with embedded subgroups.

Figures 1 and 2 show a star field in the 30 Dor region of the LMC with two different scales, from Efremov (1989). Figure 1 has a large scale and contains several groupings of stars 20′ in length in addition to much smaller sub-groupings. Figure 2 shows the part of Figure 1 that is southwest of 30 Dor (the large cluster in the center of Fig. 1). Clearly each stellar grouping contains smaller groupings in a hierarchical pattern. Very few groups are isolated, and very few contain compact stellar clusters at their cores.

### 2.1. Intrinsic correlations between physical properties: the time-size relation

Many observations have been made during the last four decades of the hierarchical structure of star-forming regions. Now we believe that the observed sequence of stellar clusterings extends from multiple stars to clusters, to associations, to groups of associations, to stellar complexes, and possibly even to short blue spiral arms, such as the Orion arm. All of these groups probably form by the same physical processes, so they are all similar in this respect. They have different nomenclature only for historical reasons, primarily because of a long history of unrecognized intrinsic correlations between other cluster properties along this sequence, properties such as density and age range. Now we suggest that all of these properties follow physically from the turbulent nature of the interstellar medium.

Along this sequence of cluster sizes, the age range between the youngest and the oldest members usually increases, implying *a longer duration for star formation inside larger groups*. A schematic representation of this is shown in Figure 3. The scaling relation between duration of star formation, $t_{SF}$, and size, $S$, has been determined from patches of Cepheid variables in the LMC and from

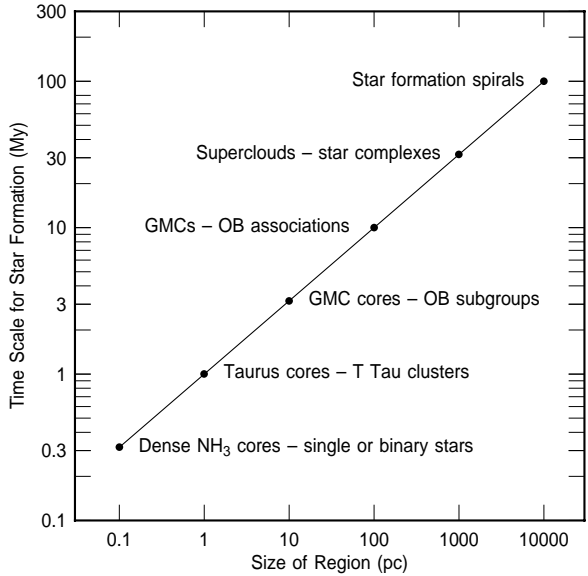

Figure 3. Schematic diagram showing how the duration of star formation in regions of various sizes increases with the square root of the size.

local star formation (Elmegreen & Efremov 1996). The same time-size relation applies in both cases, giving

$$t_{SF}(My) \approx S(pc)^{0.5}. \qquad (1)$$

This relation is similar to that obtained from the size-linewidth correlation for molecular clouds, which is shown on the top of Figure 4 using several local cloud surveys. The plotted points suggest $c(km\ s^{-1}) \sim 0.7 S(pc)^{1/2}$ for Gaussian velocity dispersion $c$ and FWHM $S$. The ratio of $S$ to $c$ is shown on the bottom of the figure. Half this ratio gives the turbulent crossing time,

$$t_{crossing}(My) \sim \frac{0.5 S(pc)}{c(km\ s^{-1})} \approx 0.7 S(pc)^{0.5}. \qquad (2)$$

The similarity between the duration-size relation for star-forming regions and the time-size relation for GMCs suggests that *the total duration of star formation in a cloud is related to turbulence* (Elmegreen & Efremov 1996).

Some of the apparent expansion of OB associations may be related to the size-velocity-time correlation. The inference that OB associations expand is based partly on the presence of runaway O-type stars like $\zeta$ Per, which presumably result from supernova explosions in binary systems. Another contributor to the expansion is from the gradual unbinding of the primordial cloud, as noted by Zwicky (1953). However, a large part of what we interpret as expansion is simply the larger sizes, and larger velocity dispersions, of older regions compared to younger regions. The usual interpretation is that the larger regions got to be so large because they expanded from smaller sizes that were similar to what we see today for the smaller regions. But this expansion need never have occurred to produce the observed effect. Stars form on all scales, with the large and small regions forming simultaneously except that the larger regions generally form stars for a longer total time than the smaller regions. Thus there are many examples of large regions (Orion OB1) containing smaller regions (the subgroups) that could have obtained this overall structure from the hierarchical nature of the primordial cloud, and not from the expansion of identical subgroups. That is, the extent of the whole Orion OB1 association is not large because most of the stars were once confined to a small volume and then expanded. It is large because the part of the primordial cloud that had a $\sim 20$ My evolution time was always large. Similarly, the subgroups are not each younger versions of the whole Orion association, ready to expand to fill the whole 100 pc of today's association, but only smaller regions of star formation inside the larger cloud, where the average gas density was always higher and the timescale for evolution always smaller. The fact that the oldest subgroup in Orion is perhaps twice as large as all of the youngest subgroup is the result of a real expansion from cloud unbinding.

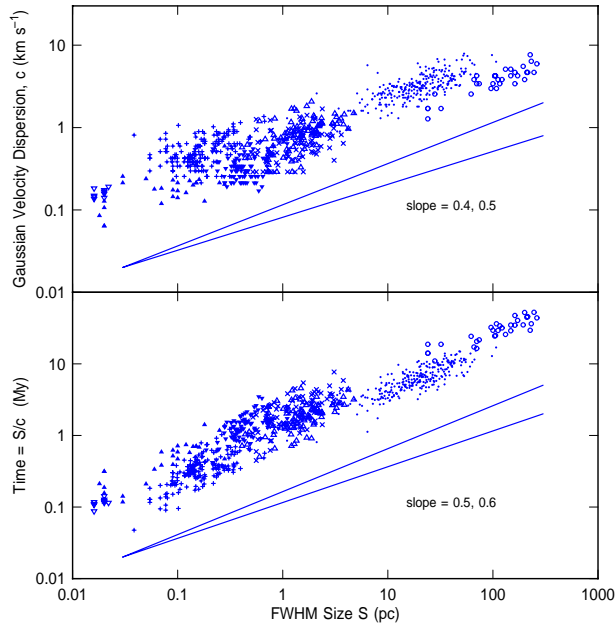

Figure 4. Size and linewidth data from GMC surveys. Symbols are for GMC surveys = *dots*: Solomon et al. (1987), *open circles*: Dame et al. (1986); quiescent clouds = *filled triangles*: Falgarone et al. (1992), *open triangles*: Williams et al. (1994; the Maddalena-Thaddeus cloud), *inverted open triangles*: Lemme et al. (1995; L1498), *inverted filled triangles*: Loren (1989; Ophiuchus), OB associations = *crosses*: Williams et al. (1994; Rosette), *plus signs*: Stutzki & Güsten (1990; M17).

The scaling factor in the size-linewidth relation for self-gravitating clouds *depends on interstellar pressure*, and can be higher in regions of higher pressure.

For virialized clouds of radius $R$, we get a velocity dispersion (Elmegreen 1989)

$$c(km\ s^{-1}) \approx 0.4 P_4^{1/4} R(pc)^{1/2}, \qquad (3)$$

where $P_4$ is the pressure in units of $10^4 k_B$ for Boltzmann constant $k_B$. This result comes from the relations $P \sim GM^2/R^4$ and $c^2 \sim 0.2GM/R$, which give $c \sim (GP)^{1/4} R^{1/2}$ after eliminating $M$. The coefficient actually includes magnetic terms in the virial theorem and a realistic density gradient in the cloud. The result compares well with the observations if we set $S = 2R$ and $P_4 \sim 40$, which is higher than the local average interstellar pressure ($P_4 \sim 3$), presumably because the overall molecular clouds are self-gravitating and they have heavy HI self-shielding layers on top of them (Elmegreen 1989).

Equation 3 suggests that the correlation between the duration of star formation and size should scale with interstellar pressure as

$$t_{SF}(My) \sim \frac{R(pc)}{c(km\ s^{-1})} \sim 2.5 P_4^{-1/4} R(pc)^{1/2}. \qquad (4)$$

Evidently, *regions with higher pressure will have faster, more intense star formation*. This is the case for starburst galaxies. The large column densities of gas, $\sigma$, in the inner regions of starburst galaxies, which are necessary to give star formation because of the high epicyclic frequencies there, make the interstellar pressure very high ($P \sim G\sigma^2$), and this makes even the large clouds form stars quickly (Elmegreen & Efremov 1997). The opposite applies to low surface brightness galaxies, which, because of their low surface densities, have very low pressures and extremely long durations of star formation for a region of any specific size. Thus starburst galaxies have spectacular bright knots of intense star formation, some containing the mass of Gould's Belt, while low-surface brightness galaxies have rather dull star formation on the same mass scale.

### 2.2. Galaxies with different sizes

The continuum of stellar aggregation extends over different galaxy sizes too. Small spiral and irregular galaxies have smaller *largest* star complexes than large galaxies, scaling as nearly the first power of the size of the galaxy, i.e., $D_{complex}(pc) \sim 0.19 D_{gal}(pc)^{0.82}$ (Elmegreen et al. 1996). Thus 30 Dor, one of the largest star-forming regions in the LMC, is quite small compared to the largest star-forming regions in giant galaxies like the Milky Way, i.e., compared to Gould's Belt. Yet because the star formation time scales with the turbulent crossing time in a region, and the velocity dispersion on a large scale in non-interacting galaxies is always about 5 to 10 km s$^{-1}$, *the largest complexes in small galaxies have smaller star formation times than the largest complexes in large galaxies*, which makes star formation seem brighter and more burst-like in small galaxies. This means that while 30 Dor is physically small compared to Gould's Belt, it is also more intense, forming all of its stars more quickly, and having a much larger density of OB stars than Gould's Belt.

The fact that the complex diameter is not exactly proportional to the galaxy size confirms the common impression that smaller galaxies have proportionately larger star-forming regions than giant galaxies. Yet this effect is rather minor: for galaxies spanning a range of a factor of $10^4$ in luminosity, the ratio of the largest complex diameter to the galaxy diameter varies by only a factor of $\sim 2$ to 4. The important point is that star forming regions are physically much smaller in smaller galaxies, and they form stars more rapidly because of their small sizes and shorter turbulent crossing times.

## 2.3. Hierarchical structure on large scales: spiral arm pieces

Another important consideration is the nature of star formation on the largest scale. How big does hierarchical structure get in a galaxy? It is fairly easy to recognize large patches of active star formation, but these patches are usually chosen to be near-circular in shape. Anything with a shape like a spiral arm would not generally be considered a coherent star formation region. But this is only a subjective decision. In fact, the largest coherent patches of star formation in galaxies should look like short spiral arms, or spurs in an otherwise grand design spiral.

For the above scaling relation between time and size, the duration of star formation in a region begins to exceed the inverse of the Oort constant $A$, which is the local shear time, when the size of the region exceeds about the disk thickness (Elmegreen & Efremov 1996). This means that regions larger than a galaxy thickness of several hundred parsecs will take so long to form stars that background galactic shear will have time to distort them into spiral-like shapes before the star formation process is over.

For a flat rotation curve, the average shape of a sheared star formation patch will be a hyperbolic spiral, and the age will be related to the spiral pitch angle $i$ by the expression

$$age = \left(-\frac{R}{\tan i}\frac{d\Omega}{dR}\right)^{-1} = \frac{1}{2A\tan i} \approx \frac{R}{V\tan i} \qquad (5)$$

for galaxy angular rotation rate $\Omega$ and galactocentric radius $R$; the last expression is for a flat rotation curve with speed $V$. This age assumes that the star formation region begins its life as a circle and simply shears into a spiral with time. Note that although the duration of star formation depends on size according to equation 1, the size does not actually appear in this expression for age when it is written in terms of the pitch angle. Local shear in a spiral density wave could change the relation between pitch angle and age. For spiral-like star formation regions with a pitch angle of $\sim 20°$ in a flat rotation curve, the age from this expression is $\sim 100$ My. Of course, spiral arm pieces like this would contain sub-condensations (star complexes) and sub-sub-condensations (OB associations), and so on, each of which might have a more globular shape because of their younger age, but together they have the appearance of a short spiral arm.

An example is in Figure 5, which illustrates the southwest quadrant of the galaxy M33. There is a long dense spiral arm that has a symmetric counterpart on the other side of the galaxy, and it also has dust lanes, indicating that it is a spiral density wave. But there are also several other arms, or arm-pieces, that are not connected to the density wave and contain only young stars. These isolated arms have no dustlanes and they have slightly larger pitch angles than the density wave. We believe that these secondary arms are pure gas and star formation, driven by self-gravity and turbulence on a large scale in the disk.

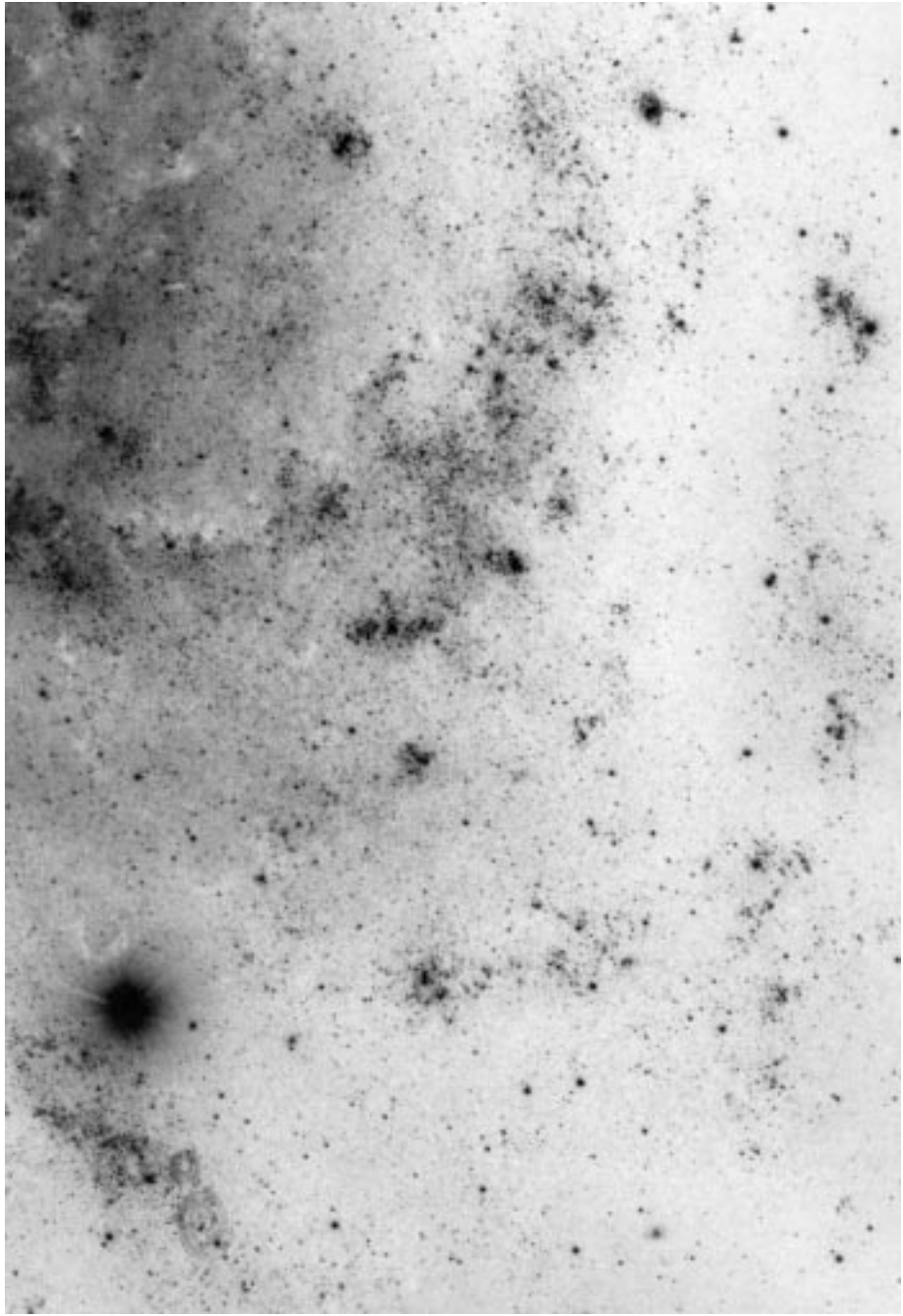

Figure 5. Southwest quadrant of M33, showing the main spiral arm in old and young stars (top center) plus several long, spiral-like groupings of young stars with hierarchical sub-clusterings (bottom right). North is up. Image from Sandage & Bedke (1988).

Note that self-gravity in the gas plays several roles in the development of large scale structure and star formation. It can make spiral arms from local irregularities that arise in random turbulent motions (the swing amplifier – Toomre 1981), and it can cause the spiraling gas to clump up into giant cloud complexes, which typically contain $10^7$ M$_\odot$ of gas (Elmegreen 1991). It can also cause the gas in spiral density waves to clump up in the same fashion (Elmegreen 1994). These giant gas clouds become subdivided further by a combination of self-gravity and turbulence compression, making the entire hierarchy of structures in essentially one crossing time. The ultimate fate of these clouds is the formation of star complexes and all of their embedded associations and sub-clusterings.

The Orion spur, sometimes called the Cygnus - Orion arm, seems to be a good example of sheared, hierarchical star formation (Elmegreen & Efremov 1996). It is not a spiral density wave because it contains *only* young star clusters (Becker 1963; Lynga 1987; Efremov 1997a). This contrasts sharply with a real density wave in our Galaxy, the Sgr-Car arm. The Sgr-Car arm clearly has both streaming motions and concentrations of old clusters and old stars (Bok 1964; Ardeberg & Maurice 1980, 1981; Avedisova 1987; Gerasimenko 1993; Efremov 1997a, 1997b), indicating that it is a dynamical feature in the old stellar disk. The Sgr-Car arm also has enhanced gas and star formation densities. However, the Ori-Cyg spur has only the enhanced star formation, suggesting it is a pure gas and star-formation feature, sheared by differential rotation.

Turbulence scaling laws make interstellar cloud structure merge with galactic spiral structure on large scales. With this in mind, the specific formation history of the Orion region may now be considered.

## 2.4. The origin of the Orion star-forming region

The Orion OB association is part of Gould's Belt and perhaps one of several gaseous condensations in the expanding Lindblad ring of HI and CO emission, along with the Sco-Cen and Perseus OB associations (Olano 1982; Elmegreen 1982). Orion and the others are less than 20 My old, while Gould's Belt is around 50 My old. Thus Gould's Belt began forming stars before the local OB associations, and when it did, it probably made the Cas-Tau, $\alpha$ Per, and Pleiades clusters, along with many dispersed B and later-type stars (Lesh 1968; Pöppel 1997).

The 50 My age of Gould's Belt is interesting because this is about the time since the local gas was inside the nearest spiral density wave arm, which is the Carina arm. The Carina arm is $\sim$ 4 kpc from the Sun along the solar circle at $l = 282°$ (Graham 1970). If the pattern speed of the wave is 13.5 km s$^{-1}$ kpc$^{-1}$ (Yuan 1969), then the physical speed of the local arm is 114 km s$^{-1}$ in the tangential direction. It follows that the time since the Carina arm passed the Sun would have been 35 My if there were no streaming motions parallel to the arm. With streaming, the time can be twice this value. This similarity with the age of Gould's Belt suggests that *Gould's Belt began as a giant gaseous condensation in the Carina spiral arm when the arm was last at the Solar position in the galaxy* (Elmegreen 1993).

Perhaps other parts of the original concentration of gas and young stars from the Carina arm are still near us, contributing to the Orion-Cygnus "arm" and other star-formation features.

Orion is not an example of the main event of star formation in galaxies. Most star formation occurs in $10^7$ M$_\odot$ concentrations of gas that form in long, density wave arms, probably as a result of gravitational instabilities that bunch up the gas along the arm (Elmegreen 1994). Orion is a minor episode in the demise of such a concentration, seen today as Gould's Belt, and which, when finished, will be a typical star complex. Thus Orion is not representative of star formation in general: galaxies do not selectively make $10^5$ M$_\odot$ clouds and OB associations like Orion. There is no characteristic scale for cloudy structure below the Galactic disk thickness.

Neither is Orion at the lower end of the hierarchy of cloud structure. It contains four smaller subgroups with star formation durations of about 5 My Blaauw (1964; 1991), and the youngest of these still show denser and smaller sub-clusters with ages much less than 1 My. Indeed, each level in Figure 3 can be found connected with the Orion region.

## 3. Hierarchical structure in star formation results from hierarchical structure in gas

### 3.1. Hierarchical structure in gas: fractal structure

Stars form in dense interstellar clouds, so the structure of young star clusters should reflect the structure of clouds. This means that if stars form in hierarchical groupings, then the dense clouds have to be hierarchical as well.

Indeed, hierarchical structure has been observed for a wide range of scales in the interstellar medium, from clumps at the resolution limit of the telescope (Gill & Henriksen 1990; Houlahan & Scalo 1992; Langer, Wilson & Anderson 1993; Pfenniger 1996) to giant spiral arm clouds (Elmegreen & Elmegreen 1983, 1987). Structure probably even exists below the resolution limit because the excitation density of many molecules exceeds the average density that the telescope sees. Furthermore, cloud structure looks about the same for distant clouds and nearby clouds at different spatial resolutions, as long as the mapped regions span a wide range of scales. This means that the mass distribution function for clumps in clouds is independent of distance. These observations imply that the hierarchical structure in clouds is approximately self-similar, in which case one can say it is *fractal* with a well-defined fractal dimension.

The fractal dimension of interstellar gas has been estimated to be $D \sim 2.3$ from wavelet analysis (Gill & Henriksen 1990) and from the size distribution of molecular clouds and clumps (Elmegreen & Falgarone 1996). This dimension appears to be independent of distance to the cloud. It is about the same as the fractal dimension of structures seen in laboratory turbulence and is therefore an indication that interstellar clouds form by processes related to turbulence. The mass spectrum of interstellar clouds and the mass-size correlation also follow from this fractal structure (Elmegreen & Falgarone 1996).

Hierarchical cloud structure with any fractal dimension has a mass distribution $n(M)dM \propto M^{-2}dM$ if all of the structure is considered with multiple counting of mass. This is because all of the mass is present at each level in the hierarchy, each level containing the other smaller levels. For example, suppose a cloud has a mass of 64 units and each level in the hierarchy has two fragments. Then there will be 1 cloud with a mass of 64, 2 clouds with masses of 32, 4 clouds with masses of 16, 8 clouds with masses of 8, 16 clouds with masses of 4, 32 clouds with masses of 2, and 64 clouds at the smallest level with masses of 1 each. These are not different clouds, but they are different structures all nested hierarchically inside each other.

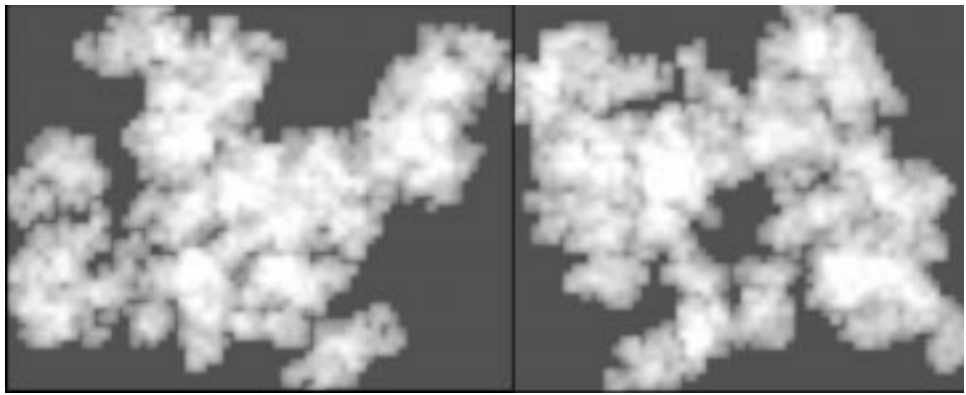

Figure 6. A fractal cloud made on a 2x2x2 nested grid with six levels is shown as front (left) and side views. The cloud is blurred to represent inadequate telescope resolution.

In this example, the logarithm of the cloud number scales linearly with the logarithm of the mass. For a log in base 2, the number of clouds as a function of $\log_2$ mass is as follows: 1 cloud, which is $2^0$ cloud, has $\log_2 M = 6$, $2^1$ clouds have $\log_2 M = 5$, $2^2$ clouds have $\log_2 M = 4$, etc., and in general, $2^h$ clouds have $\log_2 M = 6 - h$. Written in this way, the increment in mass from one level to the next is 1 in units of the log of the mass. If $\xi$ is the number of clouds in logarithmic units of mass, then $h = \log_2 \xi$ while $h = 6 - \log_2 M$. Thus $\log_2 \xi = 6 - \log_2 M$, from which we obtain $\xi(M) d \log M = 2^6 M^{-1} d \log M$. Note that the way in which the cloud is subdivided enters only into the proportionality constant, but not in the power of $M$. We can convert this into a mass distribution in linear intervals of mass by writing $n(M)dM = \xi(M)d\log M$, or $n(M) = \xi(M)d\log M/dM = \xi(M)/M$. Thus we obtain $n(M)dM = 2^6 M^{-2} dM$. *Self-similar hierarchical distributions of mass always have a mass distribution in equal intervals of mass that is $M^{-2}$, regardless of how the hierarchy is distributed.*

The fractal dimension of a hierarchical distribution is defined to equal the ratio of the logarithm of the number of pieces in a level to the logarithm of the relative size of clouds at that level. In the above example, we did not have a relative size, so the fractal dimension could be anything. If we say now that each of the two fragments in a lower level has a size that is smaller than the fragment at the next higher level by a factor of 3, then the fractal dimension is $\log 2 / \log 3 = 0.63$. This is a far more open structure than real interstellar clouds. If we consider the opposite extreme of a completely filled structure, such as tightly packed cubes within cubes, then each factor of 3 decrease in size would have a factor of $3^3 = 27$ more cubes. In this case the fractal dimension would be $\log 27 / \log 3 = 3$. This is the fractal dimension of filled space because cubes nest together to fill up space. Interstellar clouds are more open and empty than that. For a fractal dimension of 2.3, each factor of 3 decrease in scale corresponds to a factor of $\sim 12$ more fragments, because $\log 12 / \log 3 = 2.3$. Similarly, each factor of 2 decrease in scale corresponds to $\sim 5$ more fragments. Recall that Scalo (1985) noted how each level in the hierarchy of interstellar cloud structure has about 2 to 5 times more clouds. He was apparently defining the levels by nested structures that differed by factors of $\sim 2$ in scale.

Two views of the same fractal with dimension 2.3 are shown in figure 6. This was made by randomly selecting $N = 5$ subcubes inside larger cubes, with a corresponding scale shrinkage factor of $L = 2$. There are six levels total. The object is three dimensional, and viewed in projection with brightness proportional to the column density of material. The smallest scales are unresolved to make the images look more like real clouds.

The method used to derive the $M^{-2}$ mass spectrum is not the same as the method used to derive interstellar cloud or clump mass spectra. The method used above has multiple counting of the same masses, over and over again for each level, and it also ignores the interclump medium. Interstellar studies avoid this multiple counting because they want the total clump mass to add up to some fraction of the overall cloud mass, with the rest presumed to be in an interclump medium. Thus it is difficult to compare the above model to real clouds. Elmegreen & Falgarone (1996) got the fractal dimension from the *clump size distribution*, reasoning that size is a measured *geometric* property of a cloud, like a fractal dimension, whereas the mass comes from a combination of size, linewidth, and brightness temperature. They then determined the relation between size and mass empirically and converted the size spectrum to a mass spectrum (see Sect 3.3. below). The result gave the observed mass spectrum for 5 separate surveys. These mass spectra were all slightly different because the empirical conversion of size to mass was slightly different for each survey, depending on how the clumps were defined and which molecular transition was used. The important point was not the final mass spectrum, but the fact that each cloud had the same fractal dimension from the size distribution, within the errors, i.e., $2.3 \pm 0.3$, even if they had different mass spectra and distances.

The size distribution of clouds in a hierarchical distribution seems to be a good way to calculate fractal dimensions. In one of the above examples, there were 2 sub-fragments inside each fragment, and each subfragment was smaller by a factor of 3. Then there are 64 ($= 2^6$) smallest clouds of size 1, say, $2^5$ clouds of size $3^1$, $2^4$ clouds of size $3^2$, and so on up to 1 cloud of size $3^6$. For this logarithmic interval of size, the number of clouds $\xi$ varies as $2^h$ and the size $S$ varies as $S = 3^{6-h}$. Solving for $h$, and using base 10 logarithms for clarity this time, we get $h = \log \xi / \log 2$ from the number of clouds, and $h = 6 - \log S / \log 3$ from their sizes. Now we eliminate $h$ to get $\log \xi = 6 \log 2 - \log S (\log 2 / \log 3)$, or $\xi = 2^6 S^{-D}$ where $D = \log 2 / \log 3$ is the fractal dimension. As mentioned above, the general form for $D$ is

$$D = \frac{\log N}{\log L} \qquad (6)$$

where $N$ is the number of sub-fragments per fragment and $L$ is the scale factor for sub-fragments compared to fragments. This results shows how easy it is to get a fractal dimension from a compilation of cloud sizes: *the fractal dimension is the negative slope of the size distribution on a log-number versus log-size plot.* For equal intervals of size, we get

$$n(S)dS \propto S^{-D-1}dS. \tag{7}$$

Until recently, fractal structure in interstellar clouds was determined primarily from cloud perimeters, not internal clumps. Considering only the perimeter, we obtain a *projected* fractal dimension equal to the slope of the relation between the logarithm of the perimeter length versus the logarithm of the enclosed area. That is, $D_p = \log A/\log P$ for area $A$ and perimeter $P$. This method gives $D_p \sim 1.3$ for interstellar clouds (Beech 1987; Bazell & Désert 1988; Scalo 1990; Dickman, Horvath, & Margulis 1990; Falgarone, Phillips & Walker 1991; Henriksen 1991; Zimmermann & Stutzki 1992; Hetem & Lepine 1993; Vogelaar & Wakker 1994). The fractal dimension for clumping turns out to be about 1 more than the fractal dimension of the perimeter. Note that the irregular structure on the perimeter is from the same type of clumps that are viewed in projection in the interior; they would be located there if the cloud were viewed from the side.

This difference of $\sim 1$ between the fractal dimension of clumpy structure and the fractal dimension of the perimeter has not been rigorously proven for transparent density distributions, as far as we know. Beech (1992) got a result like this for crumpled paper models. Something like it is easily proven for a *slice* through a fractal cloud. Consider a model with 12 sub-fragments for each factor of 3 decrease in size. This gives $D = 2.3$ as discussed above. How many sub-fragments will appear in a slice through this cloud? The probability that the slice will go through one of the 12 sub-fragments at any particular level in the hierarchy is the ratio of the size of each sub-fragment to the fragment size, or $1/3$. This assumes isotropic clouds. Thus the number of sub-fragments that get cut by the slice equals $12/3$, or 4, and the fractal dimension of the slice is $\log 4/\log 3 = 1.3$. This is 1 less than the fractal dimension of the whole cloud. In general we can write $N$ as the number of sub-fragments per fragment and $L$ as the scale factor. Then the probability of a slice intersecting a sub-fragment in any one level is $1/L$ and the number of sub-fragments that appear in each slice is $N/L$. The fractal dimension of the slice is the logarithm of this number of sub-fragments divided by the logarithm of the scale factor, or $D_s = \log(N/L)/\log L$. But this is just $\log N/\log L - 1$, from which we obtain the relation $D_s = D - 1$ for fractal dimension $D_s$ of a slice through a fractal of dimension $D$.

The projected fractal dimension of a semi-transparent density distribution is more complicated because the sub-fragments may shield or cover each other and not be fully counted. In the limit of full covering, which is for an optically thick cloud (in terms of clump visibility, not in terms of light absorption) the fractal dimension in projection equals 2, since the projected area is fully covered with sub-fragments and there are no gaps or holes. For an optically thin cloud, in which each clump can be seen, the projected fractal dimension of the cloud interior (not the perimeter fractal dimension) is the same as the volume fractal dimension, because the total number of visible clumps is still $N$ and the scaling factor is still $L$, giving $D = \log N / \log L$ as before. If the volume fractal dimension is less than 2, then even the projected (not perimeter) fractal dimension will be less than 2, possibly equal to the volume fractal dimension. The perimeter fractal dimension is not obviously related to any of these other dimensions for a projected cloud. Nevertheless, the observations suggest that the perimeter dimension is about 1 less than the optically-thin projected fractal dimension.

We can estimate how the fractal dimension of the projection will vary as the clumps in a cloud shield each other more and more. Consider again two levels in the hierarchy with $N$ sub-fragments inside the larger fragment, and sub-fragment sizes $S_{small}$ equal to $1/L$ times the main fragment size $S_{big}$. Now define the optical depth for fragment shielding to be

$$\tau = \left(\frac{N}{S_{big}^3}\right) S_{small}^2 S_{big} \equiv \frac{N}{L^2}. \tag{8}$$

This is defined like a normal optical depth, equal to the density of sub-fragments, $N/S_{big}^3$, multiplied by the cross section of sub-fragments, $S_{small}^2$, and multiplied by the path length through the main fragment, $S_{big}$. Evidently, this is $\tau = N/L^2$. The apparent fractal dimension when some sub-fragments shield each other is the logarithm of the number of sub-fragments actually seen, divided by the logarithm of the scaling factor, and for a cloud in projection, this is

$$D \sim \frac{\log\left(L^2 \left[1 - e^{-\tau}\right]\right)}{\log L} \tag{9}$$

In the limit of large $\tau$, this equals 2 because the projected surface is completely filled, as discussed above. In the limit of small $\tau$, this equals $\log N / \log L$, which is usual definition for $D$ in an optically thin cloud.

Most interstellar clouds are optically thin in the sense that the clump density is so low at each level that there is usually only one or fewer clump per line of sight in each velocity interval. The separate velocities help to make each clump visible, so mutual shielding is not important. Perhaps there is a fundamental reason for this, but we don't know what it is. It is certainly not true for the general interstellar medium because some lines of sight to stars have severe blending of optical absorption lines.

### 3.2. Density structure and minimum clump sizes in fractal clouds

The fractal model for interstellar clouds proposes that there is a nearly self-similar hierarchical structure of clumps within clumps over a wide range of scales, from some minimum cloud size to a much larger size characteristic of the galaxy thickness, or spiral arm thickness. The gas at the minimum size is presumed to be somewhat uniform, without any further subdivision, while the gas at larger scales is presumed to be a nested collection of minimum-size pieces separated by a pervasive, low-density interclump medium, which can also be molecular in some cases. In this model, the local average density decreases with increasing scale because the overall structure becomes more and more open, with ever-enlarging gaps between clumps of minimum-size clumps. The density comes from the fractal relation between mass and size, which is $M \propto S^D$ for fractal dimension $D$. This gives an average density $<\rho> \propto S^{D-3} \sim S^{-0.7}$. This is also about what is inferred from the Larson (1981) correlations for molecular clouds.

The density at the smallest scale comes from excitation requirements for CO(2-1) and other molecules and is about $10^5$ molecules cm$^{-3}$ (Falgarone, Puget & Pérault 1992; Falgarone & Phillips 1996; Plume et al. 1997). This is also the value from thermal pressure equilibrium at a typical temperature of 10K, since the pressure in a GMC is around $10^6 k_B$ for Boltzmann constant $k_B$. Scaling up the characteristic density of $10^3$ cm$^{-3}$ at a size of 1 pc to a density of $10^5$ cm$^{-3}$, we get a size of 0.0014 pc for the smallest clumps. This corresponds to an angular size of $\sim 1''$ at a distance of 200 pc, which is about the limit of interferometric telescope resolution for CO and other molecules in nearby clouds (actually the resolution has to be lower than the cloud size by a factor of $3 - 10$ to know for sure that the cloud is not subdivided further in the same hierarchical pattern as the larger scale structure).

Such tiny features are sometimes observed optically in projection against uniform background sources, as for example in one of the globules in the globular filament GF7 (Schneider & Elmegreen 1979) that is fortuitously in front of the North America Nebula (cf. Fig. 7). The scale on the figure is such that the ends of the pointers designated $A$ and $B$ are separated by $50'$. The tiniest visible pieces in globule A are $10''$ in size (on the original Palomar plates), which corresponds to $10^{-2}$ pc for an assumed distance of 250 pc.

### 3.3. A temperature-mass correlation for fractal molecular clouds

The relatively shallow cloud and clump mass distribution functions ($\sim M^{-1.7}$) that are inferred directly from contour maps of molecular emission result in part from a correlation between brightness temperature and mass for molecular clumps, as reported in CO surveys by Loren (1989), Carr (1987), Stutzki & Güsten (1990), and Williams et al. (1994). Elmegreen & Falgarone (1996) showed that the cloud size function, $n(S)dS = S^{-D-1}dS$, which is converted into a cloud mass function, $n(M)dM = M^{-1-\alpha}dM$, using the mass-size relation $M \propto S^\kappa$, implies $\alpha = D/\kappa$. Now, cloud mass is usually defined to be proportional to $S^2 \Delta v T$ for cloud size $S$, velocity dispersion $\Delta v$, and peak brightness temperature $T$. Typically $\Delta v \propto S^{0.5}$ and $T \propto S^{0.5}$ so $M \propto S^3$ in the above clump surveys. With $D = 2.3$ from the size distribution, this implies that $\alpha = 2.3/3 \sim 0.76$, and that the cloud mass function should be $n(M)dM \propto M^{-1.76}dM$, as observed. This is distinctly different from $M^{-2}$. Note that the observation $\Delta v \propto T$ used for this derivation also implies that CO line profiles are nearly self-similar for different cloud masses and linewidths, which is also approximately true.

The origin of the temperature-mass relation is unknown. It is possible that it is simply the result of emission by unresolved and resolved clumps and by the interclump medium. The essential point is that larger regions, because of their lower average clump densities, have relatively more emission from the interclump medium. The excitation temperature of all the radiating material may be the same $\sim$ 10K or so, so the variable brightness temperature is from a variable angular filling factor of clump gas, plus a varying opacity at nearly constant filling factor of interclump gas. If the interclump gas is at too low a density to be excited to thermal temperatures, then this model requires an interclump thermal temperature that exceeds 10K.

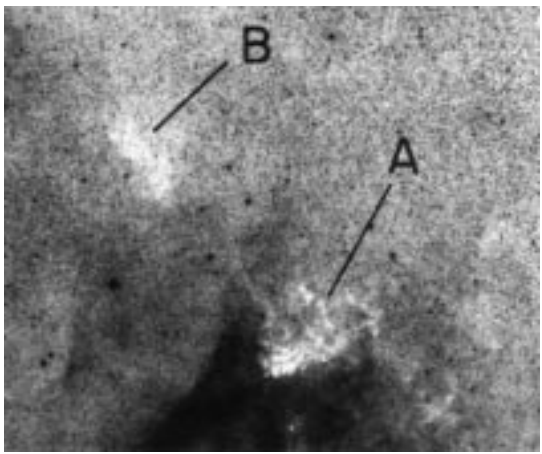

Figure 7. Globule in front of part of the North American Nebula showing small scale structure.

We can understand the two different mass spectra in the fractal model only if we consider the interclump medium, which should contribute strongly to the line profile on sufficiently large scales. The interclump medium is not observed directly, but it may be inferred to be present *and* molecular from the smoothness of the line profiles and the high opacity of total $^{12}$CO emission. In contrast, it is probably mostly atomic in the clouds of the LMC, which follows from the lower filling factor of $^{12}$CO emission in the LMC clouds (Bel, Viala, & Guidi 1986; Lequeux et al. 1994).

The angular filling factor of only the clumps in a fractal cloud *decreases* with increasing cloud size, as shown in Figure 8, which is the same as Figure 2 in Elmegreen (1997a), but with the $x$-axis reversed. This figure shows the fraction of the solid angle subtended by clumps, $f_c$, for various numbers of levels, $h$, in a hierarchical cloud made from $L \times L \times L$ cubes, for the cases $L = 3$ on the left and $L = 2$ on the right. The clump fractional solid angle decreases as the number of levels increases because a fractal cloud gets more and more open in projected solid angle as the map gets larger and larger compared to the size of the real physical matter, which is only at the smallest scale. The filling factor of real gas is nearly unity when the smallest clumps are resolved, and this limit corresponds to $f_c \sim 0.9$ at the lowest level in the hierarchy in Figure 8. As more and more levels are included, the gaps between the tiniest clumps occupy more and more solid angle, and $f_c$ decreases. Thus it follows that if a molecular cloud were only composed of clumps, the beam dilution and brightness temperature would *decrease* with larger cloud size and larger linewidth, instead of increase. An approximation to the variation in Figure 8 for the $L = 2$ case is $f_c \propto S^{-0.2}$ for $h \sim 1 - 5$.

What compensates for this decreasing clump filling factor must be an increasing relative emission from the interclump medium. This is an obvious result if the density of the interclump medium is more uniform than the spatial density of clumps. Then the ratio of the average clump to the interclump density decreases with scale approximately as the average clump density decreases, which is $<\rho>_{clumps} \propto S^{D-3} = S^{-0.7}$. At the same time, the optical depth of the interclump medium should increase with scale because this gas is volume-filling, i.e., it has a high angular filling factor on all scales, so its line-of-sight depth increases directly with $S$. This means that the optical depth for clump emission decreases with size as $\sim S^{-0.2}$, and the peak optical depth for interclump emission increases as $\sim S^{0.5}$. When the brightness temperature approximately equals the excitation temperature, the interclump medium becomes optically thick at that velocity. The square-root dependence of peak optical depth on $S$ for the interclump medium results from the fact that the total interclump column density increases linearly with $S$, but the velocity width increases with $\sim S^{0.5}$, so the peak optical depth has to increase as $S^{0.5}$ to make the product scale with the column density.

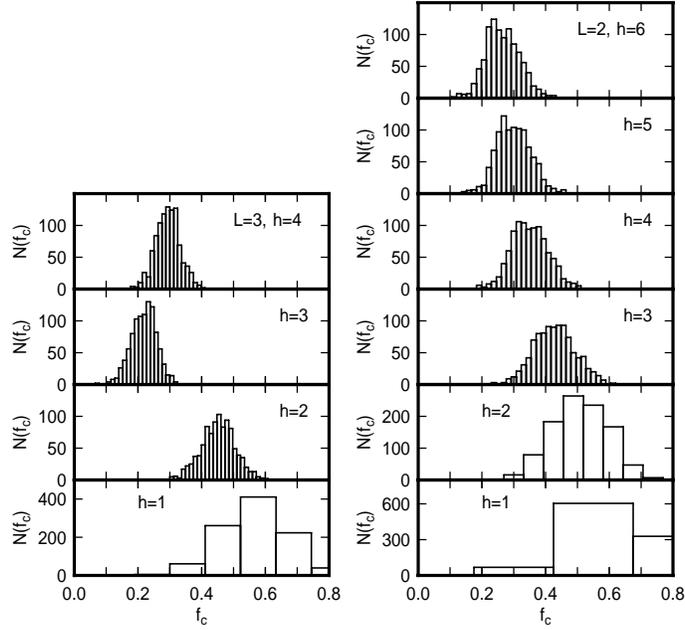

Figure 8. Fractional solid angle subtended by fragments as a function of the number of levels in the hierarchy, for a 3x3x3 nested box on the left and a 2x2x2 nested box on the right.

The net result is a trend for increasing brightness temperature with clump mass within a range of masses that exceeds the smallest clumps and still has an interclump medium that is marginally optically thin. Note that this model predicts that the brightness temperature will go up again at very small scales, when the smallest clumps (0.001 pc) are finally resolved.

This discussion assumes that the centroid velocity of the local interclump medium follows the local velocities of the clumps, which is likely if the magnetic field is well connected between the clump and interclump media (the velocity dispersion for the two media does not have to be the same on all scales, however).

We are proposing here that the difference between the $M^{-1.7}$ mass distribution function observed for clouds and clumps and the $M^{-2}$ distribution for only the clumps in a hierarchically clumped cloud is the result of a correlation between brightness temperature and mass, and that this correlation may result from an interclump medium that is distributed more uniformly than the hierarchical clumps. This simple model of brightness temperature variation with scale is shown in Figure 9. The contribution to the total brightness temperature from clumps is $T_c \left(1 - e^{-(S/S_0)^{-0.2}}\right)$, and the contribution to the brightness temperature from the interclump medium is $T_{icm}\left(1 - e^{-S^{0.5}}\right)$; the cloud and interclump brightness temperatures are $T_c$ and $T_{icm}$. This is for a fractal cloud with scales ranging from $S = 0.1 S_0$ to $S = 20$ that has a clump angular filling factor of $(1 - e^{-1})$ on the scale of $S = S_0 = 10^{-4}$ and an interclump medium optical depth of unity on the scale of $S = 1$. The plotted brightness temperature is the sum of these,

$$T_b = T_c \left(1 - e^{-(S/S_0)^{-0.2}}\right) + T_{icm}\left(1 - e^{-S^{0.5}}\right). \qquad (10)$$

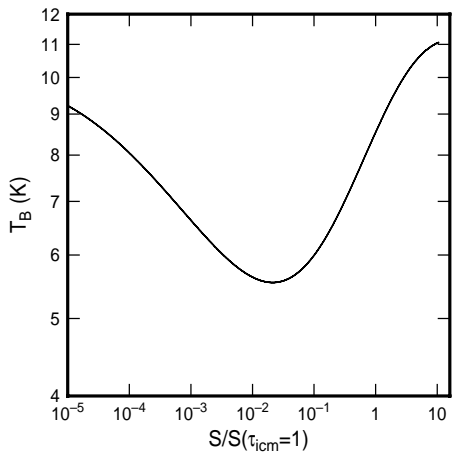

Figure 9. Model of brightness temperature versus size for fractal clumps with a uniform intercloud medium at the same temperature.

Figure 9 assumes $T_c = T_{icm} = 10 K$, although this assumption is unnecessary to get a temperature-size correlation.

This model implies that an $M^{-2}$ mass distribution should appear in surveys for which there is no temperature-size relation. An example is the Solomon et al. (1987) survey, which is temperature limited and so has a nearly constant average brightness temperature for all cloud masses. This survey also has a power in the mass-size relation that is much shallower than 3, more like $\kappa \sim D \sim 2.3$. This gives $\alpha \sim 1$ and a predicted $n(M)dM \propto M^{-2}dM$ that agrees with the observations. This is shown by the fit to the Solomon et al. observations in Elmegreen & Falgarone (1996), which is reproduced here in Figure 10. The fit actually given by Solomon et al. extends to such a low $M$ (in our opinion) that resolution limitations (for this galaxy-wide survey) have caused low mass clouds to be lost. Thus we believe Solomon et al. get too shallow a fit to the mass spectrum by extending the function to too low a mass.

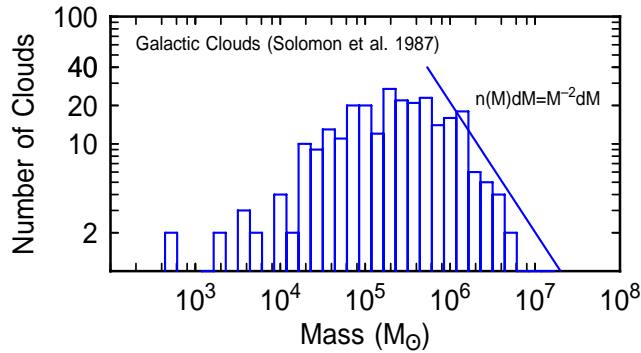

Figure 10.   Cloud mass distribution from the galactic plane survey by Solomon et al. (1987). Least squares fit to the slope of the largest clouds gives the indicated line and a mass function of $M^{-2}$.

### 3.4.   The predicted cluster mass function

A mass distribution function determined from multiple counting does not seem very interesting if observations of cloud mass distributions are done differently. But there is an important use for the $M^{-2}$ distribution that comes from hierarchical structure of clumps, and that is for the probability of selecting a single mass from the entire range of masses in the *dense*, star-forming gas. This is an important probability function for star cluster mass.

Our contention is that stars form in clusters because the clouds are clumped: star formation merely follows the dense gas. Moreover, the gas seems to have no characteristic scale once the mass is far above the thermal Jeans mass, which is typically several tenths of a solar mass. The scale-free nature may extend far below this scale too (e.g., Langer et al. 1995). Thus one can argue that there should be no characteristic scale for cluster formation either, as discussed in the first sections of this review. This property is schematically illustrated in Figure 11, which shows the same fractal as in Figure 6, but now with tiny dots representing stars at a randomly selected 5% of each position that contains mass. Clearly the dots cluster together in this figure, showing the tendency for stars to form in clusters if they follow the dense pieces of gas.

Real star formation may be more clustered than this if it requires a threshold in average density or pressure before it begins, but there is no evidence for this in purely fractal clouds; the only evidence for an *exclusion* of star formation in regions with low *average* densities is for clouds that are highly sculptured by peripheral HII regions and pressures from young stars, as is the case for the Orion molecular cloud (Lada et al. 1991). In these cases, some fractal structure could be *inside* the dense core, and subclumping of stars could occur there when they form, but larger scale clumping may not be possible because the gas has been swept into comet-like or other shapes by the older association (Bally et al. 1987).

The formation of stars within a hierarchical gas structure suggests that there should be an equal probability of forming a recognizable cluster at any level in the hierarchy, and perhaps simultaneously at many levels in the hierarchy if we look for such distributions (with a rate scaling with size, as discussed above). The point is that if the cloud properties are scale free, then any one level in the hierarchy is no more or less likely to form a cluster than any other level (once we are far from the thermal Jeans mass). In that case, the probability of forming a cluster with a mass in the range from $M$ to $M + dM$ is the same as the probability of a cloud piece having this same mass, divided by an efficiency conversion factor from gas to stars.

Generally the efficiency factor for star formation is expected to vary from region to region, but even the most extreme variation that is discussed in the literature is only a factor of $\sim 10$, from some $\sim 5\%$ efficiency to $\sim 50\%$, without any systematic trend with cloud mass. Thus, for a wide range of cloud or cluster masses, covering, say, five orders of magnitude from small globules to giant molecular clouds or giant molecular associations, the efficiency variation will not affect the slope of the cluster mass function very much in comparison to the cloud mass function. Also, if we restrict ourselves to a study of only *bound* clusters, then the efficiency of their formation is probably distributed over a much narrower range, considering that it takes a minimum efficiency of around 20% to 50% to make a bound cluster (see review in Lada 1991).

If we assume that the efficiency $\epsilon$ of star formation in a cluster is nearly constant, then the probability of forming a cluster with a mass in the range $M$ to $M + dM$ equals the probability of sampling from a hierarchical cloud a fragment with a mass in the range $M/\epsilon$ to $(M + dM)/\epsilon$. This probability is proportional to $M^{-2}$ for a hierarchical cloud, independent of the fractal dimension. Thus we expect the cluster mass function to be proportional to $\sim M^{-2}$ with these assumptions.

### 3.5. The probability of selecting a cloud fragment of mass $M$

The probability of a cloud forming a cluster with a mass between $M$ and $M + dM$ is the probability of randomly choosing a cloud fragment in this mass range. This is independent of the distribution and density of the interclump and should not mimic the observed clump mass function, which *does* include the interclump medium, as discussed above. Thus the probability for selecting a fragment can be determined from the simple fractal model.

Imagine again that a cloud is a fractal with 2 sub-fragments per fragment and any scaling of size with each hierarchy, corresponding to any fractal dimension. There are 64 pieces with mass 1, 32 pieces with mass 2, 16 with mass 4, and so on up to 1 piece with mass 64. What is the probability of selecting a piece of mass 4, say? It is the number of pieces of mass 4 divided by the total number of pieces. The number of piece of mass 4 is 16, and the total number of pieces is $1 + 2 + 4 + 8 + 16 + 32 + 64 = 127$. Thus the probability is $16/127$. Similarly the probability of selecting a piece of mass 32 is $2/127$. In general, the probability of selecting a piece of mass $M = 2^h$ is $2^{6-h}/127 \propto M^{-1}$. The same is true for other numbers of sub-fragments: the probability of selecting a certain mass is proportional to the number of fragments having that mass, and this is proportional to $M^{-1}$ for a logarithmic interval of mass, as in this example. For a linear interval in $M$, the probability is proportional to $M^{-2}$.

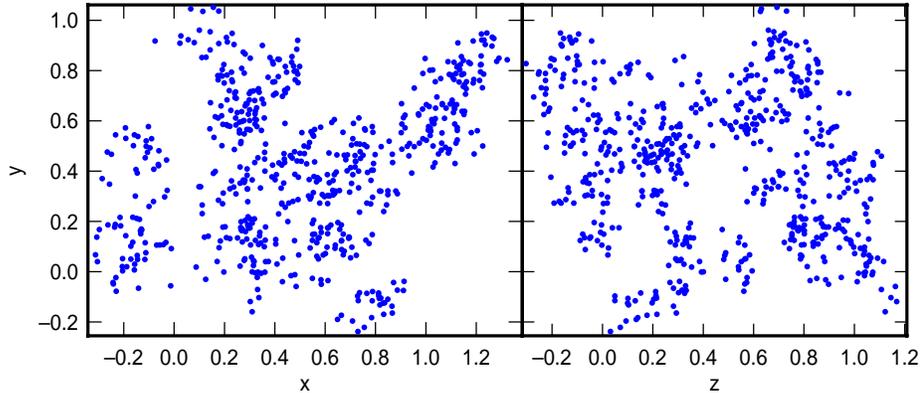

Figure 11. A fractal cloud shown previously, now with all points plotted, representing the possible distribution of stars in clumps that are unresolved in GMC surveys.

This is perhaps an obvious statement that the probability of selecting a cluster mass is proportional to the mass distribution function of the hierarchical *clumps*, not counting the interclump medium which does not form stars, except that for this probability, *it does not matter that the masses are double counted.*

Cluster luminosity distributions from the Large Magellanic Clouds are shown in Figure 12, separated into equal intervals of log age. With such a separation, the cluster luminosity is proportional to the cluster mass, so the diagrams show effectively the cluster mass function. The results indicate that once the youngest clusters, which are mostly unbound associations, are eliminated by selecting an old age, the mass function becomes $n(M)dM \propto M^{-2}dM$, as predicted for hierarchical clouds. The youngest clusters do not follow the same luminosity function because star formation is incomplete, and because the luminosity of a young cluster is strongly influenced by statistical fluctuations in a small number of O-type stars. Nevertheless, the distribution of HII region luminosities in galaxies is $L^{-2}$ also (Kennicutt, Edgar, & Hodge 1989; Comeron & Torra 1996), so OB associations in a larger sample still follow the predicted distribution. Battinelli et al. (1994) also got $M^{-2}$ mass functions for small samples of local clusters.

### 3.6. Section Summary

We conclude that stars are hierarchically grouped because of the fractal distribution of gas in interstellar clouds. There is no characteristic size for cluster formation, and the labels that have been applied to various cluster types (associations, aggregates, complexes, etc.) reflect more an observational selection of certain stars than a physical difference in cluster origin. Note that for this conclusion, we do not restrict our definition of a cluster to gravitationally bound objects.

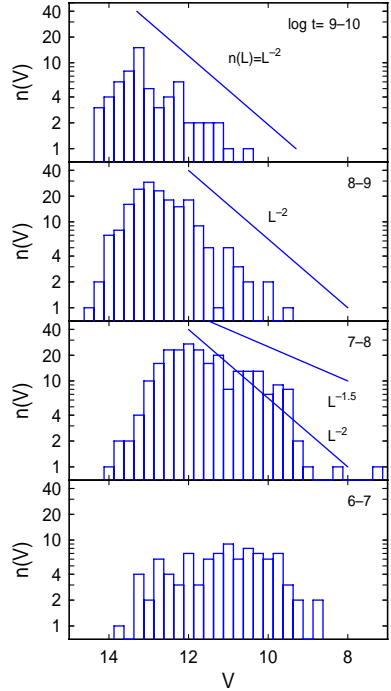

Figure 12. The luminosity distribution functions for clusters in the LMC, separated by age.

The cluster mass function is approximately $n(M)dM \propto M^{-2}dM$ because this is the probability distribution function for selecting a fragment of mass $M$ from a hierarchical distribution of fragments, regardless of the fractal dimension (which enters only into the size distribution of fragments). This cluster mass function is the same as the fragment mass function when multiple counting of fragment mass is allowed.

The *cluster* mass function is shallower than the *stellar* mass function because stars compete for gas as they form, unlike clusters. Clusters contain all of the enclosed star mass because stellar orbits mix. Even though gas that is taken first by one star goes not get into another star, both stars with this same total mass get counted in the final cluster mass. Also, the clusters that form in separate parts of a cloud do not compete for gas mass, because if they are close enough to compete for the same gas mass, then they will merge and mix into (or at least be counted as) a single cluster.

There are some possible caveats to this argument about cluster formation. The probability of forming a *bound* cluster in a normal galaxy disk environment may drop at large mass if more massive embedded clusters are proportionately more destructive of their clouds than low mass clusters, and so less likely to form stars with a net high efficiency (Elmegreen 1983). This would make the bound cluster mass function steeper than $M^{-2}$ at high cluster mass, as pointed out to B.G.E. at this conference by Dr. C. Lada. In fact, we expect this steepening to occur, but it may be so sharp that all that is noticed is an upper mass limit for bound clusters, with a sharp drop in $n(M)$ for larger masses. That is, if the steepening at high mass is sharp because the destruction of cloud fragments is sudden and efficient once the first O-type star appears, then this sharp steepening will not seriously influence the slope of the cluster mass function at lower mass. On the other hand, a loss of low mass clusters from surveys because they are too faint, or because these clusters have evaporated or been destroyed, would make the cluster mass function more shallow than $M^{-2}$. The observations of cluster masses and our understanding of possible selection effects are too primitive to say much about this at the present time.

We also wish to point out that a cluster mass function has an equal amount of star formation for each equal logarithmic interval of cluster mass. This is conceptually different from the conventional notion that most stars form in giant OB associations. In fact, most stars *do* form in OB associations, and we can also say that most stars form in star *complexes*, because of the hierarchical nature of all star formation: i.e., most of the smaller pieces are contained in the larger pieces. But if we consider the local size of a region, i.e., the local region where the star density is significantly above some average, then the $M^{-2}$ result gives an equal amount of star mass in equal logarithmic intervals of mass. This means that any particular star is equally likely to have formed in a small cluster with a mass in the range of $10^2$ to $10^3$ $M_\odot$ as in a large cluster with a mass in the range of $10^5$ to $10^6$ $M_\odot$. Nevertheless, the small cluster is probably part of the large cluster. We could be more precise saying that each particular star forms in many clusters simultaneously, since essentially all clusters are contained in other clusters. Thus the concept of the size of a region in which a particular star forms is not well defined when the hierarchical structure of clouds and clusters is considered.

## 4. Bound clusters and Globular Clusters

Although the Orion region is not forming a dense massive cluster, or globular cluster, it is nevertheless forming a large total mass of unbound stars, and it is also forming a small mass of stars in several dense clusters at the interface between the molecular cloud and the HII region. An important characteristic of this interface region is that it is at *high pressure*. Globular clusters also form at high pressure (Elmegreen & Efremov 1997), so we might learn something about cluster formation in Orion from a comparison with globular clusters.

Massive clusters that form in high pressure environments are more likely to end up bound than low mass clusters or clusters of equal mass in low pressure regions because virialized clouds are more tightly bound at high pressure. This is because the escape velocity of a cloud of mass $M$ increases with external pressure as

$$v_{escape} \sim G^{3/8} P^{1/8} M^{1/4}, \qquad (11)$$

and as the escape velocity increases, so does the difficulty of cloud disruption when star formation begins. As a result, high pressure clouds are more likely to form stars with a higher net efficiency, and therefore more likely to remain self-bound after the residual gas leaves. Thus the dense clusters that are forming at the GMC/HII region interface in Orion are more likely to end up gravitationally bound when the gas leaves than other embedded clusters of equal mass that are forming in low pressure regions.

The ambient pressure on a large scale is much lower than the pressure in the Orion ridge, perhaps by two orders of magnitude. Large scale star formation is also subject to galactic shear. This implies that whole OB associations and star complexes that form in normal galaxy disks are not likely to produce single compact clusters, like globular clusters. The O stars that form in them, combined with the galactic shear, differential flows, and variable tidal forces that occur in a spiral density wave, all cause the low pressure gas concentrations that make these stellar groupings to disperse before a high efficiency is reached.

Globular clusters formed, and are still forming, in very different environments. The oldest globular clusters in galaxies typically formed in halos before the disks were made. These halos then contained all of the galaxy's gas in a spheroidal distribution, and they had a velocity dispersion equal to the full orbit speed of the galaxy. Thus the halo pressure was higher than the current disk pressure by perhaps three orders of magnitude. Young globular clusters also form today at high pressure, in interacting or starburst galaxies where the large mass surface density of the inner disk creates a large pressure by self-gravity.

Without such large pressures, regions of star formation containing $M > 10^5$ $M_\odot$ produce only unbound associations and star complexes. An extensive discussion of this point is in Elmegreen & Efremov (1997).

## 5.  The Initial Stellar Mass Function

The power-law part of the mass distribution function for *stars* has been difficult to explain for 40 years, but there is now an explanation in terms of the fractal cloud model that has some relevance to the observations discussed in this review.

The main point of the model (Elmegreen 1997b) is that most of the stellar mass function comes from the random selection of cloud pieces, which, as we saw above, gives an $M^{-2}$ distribution. The Salpeter IMF slope is $-2.3$ in this notation, and most modern observations of the IMF in clusters and star-forming regions get the Salpeter value (see the review of observations in Elmegreen 1997b). Thus we have to explain how a fragment selection probability that scales with $M^{-2}$ changes to a stellar mass function that scales with $M^{-2.3}$.

First of all, this difference in powers could be entirely the result of a varying fraction of fragment mass that gets into the star, but this is *ad hoc*, and there are better explanations that seem more likely.

A better explanation is that the *selection* of cloud pieces to make a star, while random, will still occur in some order, and this order is likely to be strongly influenced by local average density. This is because both the dynamical time scale and the magnetic diffusion time scale vary with the inverse square root of the local average density. Thus stars should form faster in clumps with higher densities. It then follows, that because the local average density in a hierarchical cloud increases at low mass, the selection of low mass fragments to make a star will be preferred, in the sense that these masses will be selected more often and selected first. In fact, the inverse square root of density varies approximately as $M^{-0.15}$, so there is this additional amount of steepening in the mass function from a density dependence.

But density variations are not the complete story. Stars also compete for mass, so the first star to get a piece of gas makes it unavailable to other stars. This competition makes the spectrum slightly steeper still, with a resulting slope that is almost exactly $-2.3$, independent of fractal dimension or any cloud properties.

Now it is evident why the stellar mass function is steeper than the cluster mass function: clusters do not compete for mass, the different pieces of a cloud that make cluster stars simply merge into a larger cluster as the stellar orbits mix. If two clusters that form in the same cloud are so far apart that they do not merge, then they do not compete for gas mass either. The formation order of clusters does not matter for the final cluster mass spectrum if they do not complete for mass. Neither does their formation time, because once a cluster is completely made, the time it took to do so is unimportant. For stars, the formation time matters because of the competition for mass. Thus stars pick up an extra power of $M^{-0.3}$ in their mass function, whereas clusters do not. Both are steeper than the observed *cloud* mass function, perhaps for the reasons given in Section 3.3., which has to do with the presence of an intercloud medium that is unrelated to star formation.